\documentclass{iopart}

\usepackage{graphicx}
\usepackage{amsfonts}

\begin{document}

\newcommand{\myL}{\mathcal{L}}
\newcommand{\del}{\nabla}
\newcommand{\Real}{\mathbb{R}}
\newcommand{\interp}{I_{2h}^h\,}
\newcommand{\prolong}{I_{2h}^h\,}
\newcommand{\restrict}{I_{h}^{2h}\,}
\newcommand{\mkeq}[1]{\begin{equation}#1\end{equation}}

\title[Multigrid methods on domains containing holes]
     {Tips for implementing multigrid methods on domains containing holes}

\author{Scott H. Hawley and Richard A. Matzner}

\address{Center for Relativity,
         University of Texas at Austin,
         Austin, Texas 78712 USA} 
\ead{shawley@physics.utexas.edu, matzner@physics.utexas.edu}

\pacs{0.420C, 0.425, 0.425Dm}

\begin{abstract}
As part of our development of a computer code to perform 3D
`constrained evolution' of Einstein's equations in 3+1 form, we
discuss issues regarding the efficient solution of elliptic equations
on domains containing holes (i.e., excised regions), via the multigrid
method.  We consider as a test case the Poisson equation with a
nonlinear term added, as a means of illustrating the principles
involved, and move to a ``real world" 3-dimensional problem which
is the solution of the conformally flat Hamiltonian constraint with 
Dirichlet
and Robin boundary conditions.  Using our vertex-centered multigrid
code, we demonstrate globally second-order-accurate solutions of
elliptic equations over domains containing holes, in two and three
spatial dimensions.  Keys to the success of this method are the
choice of the restriction operator near the holes and definition
of the location of the inner boundary.  In some cases (e.g. two
holes in two dimensions), more and more smoothing may be required
as the mesh spacing decreases to zero; however for the resolutions
currently of interest to many numerical relativists, it is feasible
to maintain second order convergence by concentrating smoothing
(spatially) where it is needed most.  This paper, and our publicly
available source code, are intended to serve as semi-pedagogical
guides for those who may wish to implement similar schemes.
\end{abstract}


\section{Introduction}
Solving Einstein's equation in 3+1 form \cite{ADM,York} requires
that a set of elliptic (or quasi-elliptic) constraint equations be
satisfied at all times.  The Bianchi identities ensure that, given
a set of initial data which analytically satisfy the constraints,
the subsequent analytically evolved variables will also satisfy the
constraints.   In numerical solutions of Einstein's equations,
however, the constraints are not preserved exactly.  Thus errors
will arise as the simulation proceeds, and the extent to which the
numerical solutions actually reflect the true solutions of the
analytic equations is still an open area of research (e.g.,
\cite{Siebel,Calabrese,Tiglio,Yoneda,Apples}).  Apart from the issue
of the accuracy of the solutions obtained, there is also the problem
of numerical (in)stability, which seems to be related to the lack
of preservation of the constraints.  

It has been observed for lower-dimensional computational relativity
simulations \cite{Abrahams, Mattconstrained, Mattconstrained2} that
the use of ``constrained evolution'' schemes, in which the constraint
equations are satisfied (to some fixed accuracy) at all timesteps,
can offer better stability behavior.  Such schemes have not yet
received adequate attention in 3D simulations simply because of the
costly nature of solving the constraints at every timestep.  We are
among a set of researchers (e.g., \cite{Erik,DaveMeier,AndersonMatzner})
interested in exploring the benefits of constrained evolution for
3D applications.

Our desire to solve the constraints at each timestep motivates us
to develop a {\em fast} elliptic solver with which
to enforce the constraints, and we turn our attention to one of
the most popular methods currently in use: the multigrid method
\cite{ABrandt}.   The multigrid method is particularly attractive
because it is an optimal method, i.e. it requires only $O(N)$ 
operations, where $N$ is the number of unknowns.  Furthermore,
the multigrid method is not especially difficult to implement. 
(Although multidomain pseudospectral collocation
methods \cite{Kidder,Pfeiffer} also offer very efficient solutions
of elliptic problems, we focus on multigrid because of the relative
ease with which one can develop a multigrid code.  It is not
our intent in this paper to present a comparison between multigrid
methods and other numerical methods.) In the course
of developing our own implementation of the multigrid method
containing holes (i.e., excised regions), it came to our attention
that some researchers consider it impossible, or at least infeasible
in practice, to obtain generic results which have everywhere the
same order of accuracy as the finite difference scheme employed,
for domains containing holes, particularly in three dimensions.
(Specifically, that, using second-order-accurate difference stencils,
one could not generically obtain second-order-accurate multigrid
solutions if part of the domain was excised.)

Since we expect to use exact, analytic data as an inner
Dirichlet boundary condition (at the surface of the hole), we
believe it is not necessary to employ `sophisticated' schemes
such as deferred correction \cite{Klasky}, which can be used to 
achieve higher order accuracy for various boundary conditions.  
We are already aware that one can fairly easily
obtain second order convergence using square (in 2D) or cubical
(in 3D) excision regions in which the physical domain of the hole
is the same on all multigrid levels. However, we wish to apply
our multigrid code in situations where the shapes of the holes are
more general (e.g., spherical holes).  Thus we are motivated to find
a simple, robust method to handle the physical situations of interest
to us.

If the multigrid solver is to be used to provide initial data for
a numerical evolution code, the error of the initial solution need
only be below the truncation error of the finite difference scheme
used for evolution \cite{Mattcomment}.  Thus our pursuit of
specifically {\it second order} convergence for the multigrid scheme
might be regarded as unnecessary.  Our motivation is partly to
develop a simple algorithm which nevertheless offers fast (better
than first order) convergence, and partly because, given our exact
Dirichlet conditions for the inner boundary, we simply expect that
it should not be difficult to obtain second order convergence from
the multigrid solver.


Although it is likely the case that some of the techniques presented
here are known to some specialists, we believe it is worthwhile to 
popularize the straightforward methods we have used.

If we were applying a numerical method to a new physical system,
we would want to provide a full discussion of mathematical foundations,
for issues such as existence and uniqueness of the solutions.  Since
the use of finite difference techniques to solve the constraint
equations is by now a standard activity in relativity, and the
properties of the constraint equations are well known (for the
Euclidean, constant-mean-curvature background and for the signs of
the solutions we use in this paper \cite{Choquet,Dain2001,Maxwell}),
we will focus our attention on the benefits of our multigrid
implementation: second order accuracy near the hole via a simple
method, and the efficiency improvements afforded by concentrating
smoothing near the hole and performing more smoothing sweeps during
the pre-Coarse-Grid-Correction (CGC) phase than during the post-CGC
phase.   The validity of our results will be established by
demonstrating second order convergence to known exact solutions as
well as second order behaviour in the truncation error.

The organization of this paper is as follows: We begin with an
overview of the multigrid method.  We then discuss some results
obtained in two dimensions, which we then extend toward a more
difficult equation in three dimensions, eventually demonstrating
the method by solving the Hamiltonian constraint equation for a black
hole (with additional matter) on a Euclidean background.  Finally
we share some ideas for more general applications, and we end with
a summary of the main five `tips' we have for others wishing to
implement multigrid methods on domains with holes.

\section{Overview of Multigrid}
The multigrid method, first introduced by Brandt \cite{ABrandt},
has received considerable attention in the literature, and is the
subject of numerous articles, conferences, reviews and books (e.g.,
\cite{HackTrot,Hackbusch,Stueben,Wesseling,WesselingBook,Briggs}).
However, its application on domains with holes --- or on domains
with ``irregular boundaries'' in general --- has received only
modest attention.  The excellent works of Johansen and Colella
\cite{Johansen} and Udaykumar et al. \cite{Udaykumar} for cell-centered
multigrid are notable exceptions.  The ``BAM'' code 
\cite{BAM} is a notable introduction of the multigrid method in
numerical relativity, and even features the ability to handle domains
with multiples holes.  (It does not employ the advances we describe
in this paper, notably our method of obtaining second order accuracy
near the hole.)  Despite this significant result, multigrid methods
in numerical relativity remain as tools of only a few specialists.
We thus thought it worthwhile to share our experiences and methods
with {\em vertex-centered} multigrid with the numerical relativity
community.

A simple algorithm to solve discretized elliptic equations is
Gauss-Seidel relaxation \cite{NumRecipes}. This method works very
poorly at high resolutions because it fails to operate efficiently
on long-wavelength components of the error. However, it is extremely
effective at eliminating short-wavelength components of the error,
or in other words, at ``smoothing'' the error (i.e.,  the residual,
see below).  The multigrid scheme is essentially a clever means of
eliminating successive wavelength-components of the error via the
use of relaxation at multiple spatial scales.  

Here we give a very brief overview of the multigrid method, following
the notes by Choptuik \cite{Mattsnotes}.  (Introductions to 
multigrid
applications in numerical relativity are also found in Choptuik and Unruh
\cite{ChopUnruh} and Brandt \cite{BrandtSchw}.) We want to solve
a continuum differential equation $\myL u = f$, where $\myL$ is a
differential operator, $f$ is some right hand side, and $u$ is the
solution we wish to obtain.  We discretize this differential
equation into a {\it difference} equation on some grid (or lattice) with
uniform spacing $h$:
\begin{equation} 
   \myL^h u^h = f^h, 
\label{differenceeq}
\end{equation}
where $u^h$ is the {\em exact} solution of this discrete equation,
and $\lim_{h\rightarrow 0} u^h = u$.
Rather than attempting to solve (\ref{differenceeq}) directly via the
costly operation of matrix inversion, we apply an iterative solution
method.  At any step in our iteration, we will have only an
approximate solution
$\tilde{u}^h \simeq u^h$, such 
that \mkeq{ \myL^h \tilde{u}^h - f^h = \tilde{r}^h, }
where $\tilde{r}^h$ is some small quantity called the {\it residual}.
The variables $u^h$ and $\tilde{u}^h$ are related by
\mkeq{ u^h = \tilde{u}^h + v^h, \label{u,tildeu,v} }
where $v^h$ is some {\em correction} term.
In this iterative algorithm, we start with some guess 
$\tilde{u}^h_{\rm old}$
and try to bring it closer to $u^h$ by applying some (approximate) 
correction:
\mkeq{ \tilde{u}^h_{\rm new} := \tilde{u}^h_{\rm old} + \tilde{v}^h
\label{uneweq}}
The remainder of the scheme involves making a choice of what to 
use for $\tilde{v}^h$.  A simple choice, called the Linear 
Correction Scheme (LCS), can be used for linear
operators $\myL^h$. 
This method can then be extended to work with nonlinear
operators via the Full Approximation Storage (FAS) method, below.

\subsection{Linear Correction Scheme (LCS)}
For linear $\myL$, we begin by
substituting (\ref{u,tildeu,v}) into (\ref{differenceeq}):
\mkeq{ \myL^h ( \tilde{u}^h + v^h ) = f^h. \nonumber}
Then we use the linearity of $\myL$:
\mkeq{ \myL^h \tilde{u}^h + \myL^h v^h  = f^h, \nonumber}
\begin{eqnarray}
\myL^h v^h  &=& f^h -  \myL^h \tilde{u}^h \cr
    \mbox{} &=& - \tilde{r}^h. 
\label{introresid}
\end{eqnarray}

Consider a coarser-grid version of (\ref{introresid}), in which
we use a mesh spacing of $2h$:
\mkeq{ \myL^{2h} v^{2h} = - \tilde{r}^{2h}. 
\label{r2heq}}

Now we come to one of the `tricks' of the multigrid method: 
Instead of the coarser residual in equation (\ref{r2heq}), 
use 
\mkeq{ \tilde{r}^{2h} = \restrict\tilde{r}^h,} 
where
$\restrict$ is a {\it restriction} operator, which 
maps values from the fine grid to the coarse grid via some 
weighted averaging operation. So then we have
\mkeq{ \myL^{2h} \tilde{v}^{2h} = - \restrict \tilde{r}^{h}.
 \label{LCSCGeq}}
(Note the tilde on $\tilde{v}^{2h}$, denoting that this is not the
exact $v^{2h}$, because we
use a residual restricted from the fine grid.)
Equation (\ref{LCSCGeq}) can be solved 
``exactly'' for $\tilde{v}^{2h}$ because this is inexpensive to do
on the coarse grid.

A second `trick' is that, for our correction term $\tilde{v}^h$ in our 
numerical update scheme (\ref{uneweq}), we 
do not directly solve equation (\ref{introresid}), but instead
we use  $\tilde{v}^h$ approximated from the coarser ($2h$) correction
$\tilde{v}^{2h}$: 
\mkeq{ \tilde{v}^h = \interp \tilde{v}^{2h},}
where $\interp$ is an interpolation or {\it prolongation} operator,
which maps values from the coarse grid to the fine grid via some
interpolation operation.
Thus we use $\tilde{v}^{2h}$ as a {\it coarse grid correction} (CGC)
to $\tilde{u}^h$:

\mkeq{ \tilde{u}^h_{\rm new} := \tilde{u}^h_{\rm old} + \interp \tilde{v}^{2h}. \label{linearCGC}}

In performing the restriction $\tilde{r}^{2h} = \restrict\tilde{r}^h$,
we are assuming that $\tilde{r}^h$ is sufficiently smooth to be
sensibly represented on the coarse grid (e.g., without aliasing effects).
This implies that $\tilde{u}^h$ also needs to be smooth for this
restriction to produce meaningful results.  Therefore, before each
restriction operation, we apply a series of ``smoothing sweeps''
to $\tilde{u}^h$ in an effort to smooth the residual $\tilde{r}^h$,
using the efficient smoothing algorithm of Gauss-Seidel
relaxation.

\subsection{Full Approximation Storage (FAS) method}
For nonlinear operators $\myL^h$, we must modify the algorithm outlined
above.
Our implementation relies on the so-called ``alternative'' 
description of the FAS algorithm, which involves a
notion of the truncation error $\tau^{2h}$, 
defined on the coarse grid by 
\mkeq{ \tau^{2h} \equiv \myL^{2h} u - f^{2h}, }
where $u$ is the exact solution to the continuum equation.   
The function $\tau^{2h}$ 
can be regarded as a correction term which makes
the finite difference equation produce the continuum solution:
\mkeq{ \myL^{2h} u = f^{2h} + \tau^{2h}. \label{taucgeq}}
For general problems, the continuum solution $u$ and truncation
error $\tau^{2h}$ may not be available, however we may use an 
approximation to $\tau^{2h}$ which is the
{\em relative} truncation error between the coarse and fine grids, 
$\tau_h^{2h}$, given by
\mkeq{ \tau_h^{2h} \equiv 
       \myL^{2h} \restrict \tilde{u}^h - \restrict \myL^h \tilde{u}^h
. \label{tauh2hdef}}
Using this in (\ref{taucgeq}) gives
\mkeq{  \myL^{2h} u^{2h} \simeq f^{2h} + \tau_h^{2h}, \label{FASCGeq}}
and it is this equation which we solve on the coarse grid.

To obtain the CGC to a fine grid solution,
we substitute (\ref{tauh2hdef}) into (\ref{FASCGeq}), obtaining
\mkeq{ \myL^{2h} u^{2h} - \myL^{2h} \restrict \tilde{u}^h 
         = \restrict(f^h - \myL^h \tilde{u}^h),}
where the term $\restrict f^h$ on the right is introduced for $f^{2h}$,
in analogy with equation (\ref{r2heq}).
By analogy to (\ref{LCSCGeq}) in the LCS scheme, the term we should use 
for the CGC is (the part on the left hand side of the
previous equation, without the $\myL$'s):
\mkeq{ \tilde{v}^{2h}  = u^{2h} - \restrict \tilde{u}^h.\nonumber}

Thus we arrive an update scheme of
\mkeq{ \tilde{u}^h_{\rm new} := \tilde{u}^h_{\rm old} + 
   \interp (u^{2h} - \restrict \tilde{u}^{h}).}
Note that $u^{2h}$ is the exact solution to (\ref{FASCGeq}), 
and can be obtained with little
effort due to the low resolution on the coarse grid.
(If there are more than two multigrid levels involved in the solution,
then $u^{2h}$ should be the best approximation $\tilde{u}^{2h}$ obtained
on the coarser grid.)

\subsection{V-Cycles and the Full Multigrid Algorithm}
Instead of only using two grids as described above, one could find
the coarse grid correction (CGC) to a fine-grid-problem by solving
for a CGC from an even coarser grid, i.e.,  obtain $\tilde{u}^{2h}
\simeq u^{2h}$ by finding $u^{4h}$
(in which case the corresponding right hand side is
$f^{4h}+\tau_{2h}^{4h}$, where 
$f^{4h} \equiv I_{2h}^{4h}[ f^{2h} + \tau_h^{2h}]$ ).  
One can imagine a hierarchy
of multiple such grids, in which coarser grids provide CGCs for
finer grids.  The solution algorithm will then take the form of
a {\em V-cycle}, in which we start with an initial
guess on the fine grid, at multigrid level $l_{\rm max}$.  Then we
perform some number of smoothing sweeps and restrict the data to
a coarser grid.  We continue smoothing and restricting to coarser
grids until we arrive at a grid coarse enough to solve the coarse
grid equation (\ref{FASCGeq}) `exactly' (i.e., to machine precision), 
at minimal computational
cost.  This coarsest grid is at level $l_{\rm min}$.  We then
prolongate this solution to finer grids by performing a series of
coarse-grid corrections, with perhaps additional smoothing operations
being performed before moving to each finer grid.

In addition, before starting a V-cycle from the finest grid, we can
use an initial guess obtained from a prior solution at a coarser
resolution.  Doing this for each grid level results in the
{\it Full Multigrid Algorithm} (FMA).   We outline the FMA as follows,
(using a notation where superscripts refer to multigrid levels,
with $l_{\rm min}$ the coarsest level, and $l_{\rm max}$ the finest):
\hfil\break
\hfil\break
\begin{tabular}{ll}
$u^{l_{\rm min}}=u_0$    & // Initial guess, e.g. $u_0=0$ \\
                         & // or $u_0=1$  \\
Solve $L^{l_{\rm min}} u^{l_{\rm min}} = f^{l_{\rm min}}$ `exactly' &  \\
do $l$ = $l_{\rm min}+1$ to $l_{\rm max}$  &     \\
\indent   $u^l = I_{l-1}^l u^{l-1}$  & // Initial guess for fine grid 
                                \\
\indent    $\mbox{}$  & // Begin V-cycle \\
\indent   do $m$ = $l$ to $l_{\rm min}+1$  &  \\
\indent\indent      Smooth (solve via Gauss-Seidel) at level $m$  & // Pre-CGC
      smooths \\
\indent\indent      $f^{m-1} = I_m^{m-1} f^m + \tau_m^{m-1}$  &
                    // Restrict to coarser grid\\
\indent\indent      $\tilde{u}^{m-1} = I_m^{m-1} \tilde{u}^m$ & 
                    // Restrict to coarser grid\\
\indent    end do  & \\
  & \\
\indent    Solve $L^{l_{\rm min}}\, u^{l_{\rm min}} = f^{l_{\rm min}}$ `exactly'
  & // Solve on coarsest grid \\
  & \\
\indent   do $m$ = $l_{\rm min}+1$ to $l_{\rm max}$ &  \\
\indent\indent      $u^l := u^l + I_{m-1}^m \left( u^{m-1} - I_m^{m-1} u^m \right)$ & // Apply CGC \\
\indent\indent      Smooth at level $m-1$ & // Post-CGC smooths \\
\indent    end do  & \\
\indent    $\mbox{}$  & // End V-cycle \\
end do   & // End FMA 
\end{tabular}

Thus we see that on all grids except the coarsest grid, we only
smooth the error, and we solve the difference equation exactly 
only on the coarsest grid.


\section{Solution of a Nonlinear Poisson Equation}
Since we are interested in ultimately solving for the constraint
equations in general relativity, we consider the Hamiltonian constraint
in a conformally flat background geometry.  This yields the
the Poisson equation with a pair of nonlinear terms added:
\mkeq{
 \nabla^2 u(x,y,z) - K^2 u^5(x,y,z) + {A^2 \over u^7(x,y,z)} = f(x,y,z),
\label{ConstrEq}
}
where $\nabla^2$ is the usual Laplacian in Euclidean space.  $K^2$
and $A^2$ are arbitrary positive real constants related to the rate
of expansion for the 3-space for which is (\ref{ConstrEq}) is the
constraint equation.  $f(x,y,z)$ is related to the energy density
in the 3-space, and can be chosen such that the resulting $u(x,y,z)$
has some known (exact) form by which we can check our numerical
results.

Rather than begin with this somewhat complicated equation,
we begin by solving
a slightly simpler-looking equation in two and three dimensions, in
which the nonlinearity is of a quadratic form, as in
\mkeq{
\del^2 u(x,y,z) + \sigma u^2(x,y,z) = f(x,y,z), 
\label{NLPoisson} } 
where $\sigma$ ($\in \Real$) is some tunable parameter 
which we typically set to $\pm 1$ (although there are good reasons
to prefer $-1$ over $+1$; see footnote regarding smoothing operations,
on the next page).  We chose this equation not because of any particular
physical relevance, but simply because it was
the equation used in the version of Choptuik's 2D multigrid code 
\cite{Mattscode}, with which we had prior experience.

Before proceeding to 3D calculations, we will describe our
implementation for solving a 2D version of (\ref{NLPoisson}).  
Due to memory limitations (and array-indexing limitations
in many Fortran compilers), we can run our (non-parallel) 3D code
only at much lower resolutions than we are able to achieve with
the 2D code.  Since some the interesting features in the convergence
studies appear only only at very high resolutions in our 2D results,
we first present our 2D scheme and the results yielded by it.

\subsection{Two dimensions}
Our starting point is a 2D FAS multigrid solver by Choptuik 
\cite{Mattscode}, which solves the equation
\mkeq { {\partial^2 \over \partial x^2} u(x,y) +
        {\partial^2 \over \partial y^2} u(x,y) 
        + \sigma u^2(x,y) = f(x,y), 
\label{2dPoisson} 
} 
on a domain $\Omega$ with coordinate 
ranges $[0,0]$ to $[1,1]$, and
subject to Dirichlet conditions at the outer boundary:
$u(x,y)|_{d\Omega_O} = const.$, and we choose this constant
to be zero.
The function $f(x,y)$ is chosen such that the solution is
\mkeq{ u(x,y) = \sin(\pi l_x x) \sin(\pi l_y y), 
\label{2dSine}}
where $l_x$ and $l_y$ are integers (which we set to unity for the
cases presented in this paper).
We also choose $\sigma=1$.
Choptuik's form assumed a domain without holes, but we extend this to 
configurations with holes, and on these inner boundaries we also 
apply Dirichlet conditions, the values
for which are discussed below.

The code uses a hierarchy of so-called ``vertex centered'' grids.
Each grid at multigrid level $l$ is a square lattice having $2^l +
1$ grid points along each edge.  The grids have uniform spacing
$h_l = 2^{-l}$ in both $x$ and $y$ directions, and the grid points
are denoted with indices $i$ and $j$ in the $x$ and $y$ directions,
respectively, e.g.,  $u(ih_l,jh_l) \simeq \tilde{u}^h_{i,j}$.

To Choptuik's original code, we added added features needed to
handle the excision region.  We also performed a few minor
optimizations.  At each grid point, we also store an integer code
which denotes whether the point is a boundary point, an excised
point, or an interior (or {\em active}) point.  The set of all excised
points is known as the {\em excision mask} or {\em mask}, and for
the cases considered in this paper, we define the mask to be those
points which are a distance $r \le r_{\rm mask}$ from the center
of the grid, where $r_{\rm mask}$ is some value of our own choosing.
The outermost points of the mask are where the inner boundary
conditions are applied; we refer to these points as ``inner boundary
points.'' In addition to the single-hole case, we will consider the
case of two holes, for which the excision regions will be points
within the radius $r_{\rm mask}$ from the center of each hole.

\hfil\break
\noindent{\em Boundary Conditions}\hfil\break

We use Dirichlet boundary conditions at both the inner boundary
(edge of the hole) and the outer boundary, and apply them on all
grid levels.   At the outer boundary, we simply set $\tilde{u}^h=0$.
At the inner boundary, we use the values of the exact solution,
i.e., we set $\tilde{u}^h=u(x,y).$

\begin{figure}
\centering
\includegraphics[width=4in]{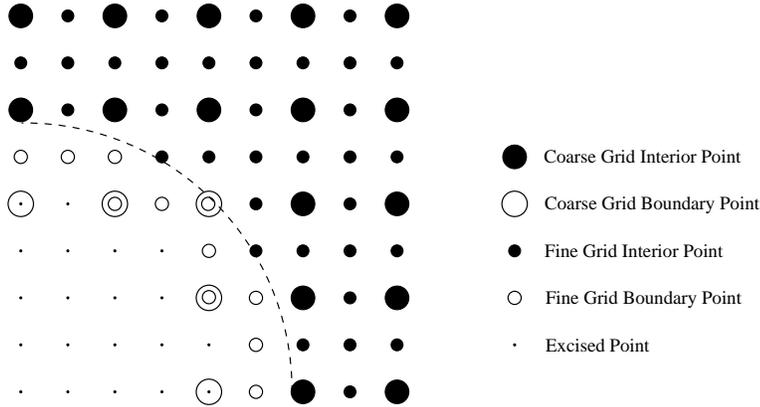}
\caption{
Example of how the inner boundary is defined, showing points on a
coarse grid and a fine grid.
Inner boundary points are those points which are immediately
interior to a circle of radius $r_{\rm max}$.
The large filled circles show normal interior grid points (i.e.,
non-excised, non-boundary points) on the
coarse grid, and the large open circles show boundary points on
the coarse grid.  The small filled and open circles show fine grid
interior points and boundary points, respectively.
The small dots show excised points on the fine and/or coarse grids, as 
appropriate.
(Only one quadrant of a full
domain is shown in this picture for purposes of clarity, but our
simulations are usually performed with the circle being centered
on the grid.)
}
\label{fig_define_ex_reg}
\end{figure}

The inner boundary is given by the outer edge of the points comprising
the excision mask; an example is shown in Figure \ref{fig_define_ex_reg}.
Thus, in some sense these boundary points are not truly ``excised''
since they have data on them.  This has the consequence that the
``size'' of the excision region (i.e., the area of the convex hull
of the data points comprising the mask) on finer grids is always
equal to or greater than the size of the excision region on coarser
grids.  This choice of the inner boundary, as being those points
just inside some level surface instead of just outside it, is
somewhat at variance from other excision schemes used in numerical
relativity (e.g.,  \cite{Frans}).  Our choice is not fundamentally
based on physical or mathematical principles, and is thus somewhat
arbitrary. However we find that this definition of the inner boundary
appears to be a key to our results of second order convergence:
if we instead define inner boundary points to be those just 
{\em beyond} $r=r_{\rm max}$, 
then we do not obtain second order convergence.

\hfil\break
\noindent{\em Smoothing Operations}\hfil\break

We use a typical ``red-black'' 
Gauss-Seidel Newton iteration to smooth the error on each 
grid, in which we loop over all `interior' points (i.e. non-excised,
non-boundary points) and apply the following two 
equations
\footnote{
We note an example of the care that must be taken in any solution
of a discretized problem. Convergence of the solution may be
impossible to achieve; this may signal significant analytical
defects in the formulation.  If $\tilde u >0$, then  $\sigma=-1$ 
guarantees that the Jacobian $2\sigma \tilde u_{i,j} - 4/h^2$ 
is nonzero (its
inverse is nonsingular in Eq. (\ref{2dgaussSeidel})). 
Clearly, there are in principle situations with 
$\sigma \tilde u> 0$ where the Jacobian can vanish. Further, because the
equations we wish to solve are nonlinear, zeros of the Jacobian
depend on $\tilde u$ and may occur at isolated spatial points.
Further, because  $2\sigma \tilde u_{i,j} $ is similar on all grids,
but $-1/h^2$ is resolution dependent, this problem can appear
differently on different multigrid levels.  It may be that a more
sophisticated solver could integrate through those singular points,
but we do not address that question here.  Notice that because 
$\tilde u_{i,j}$ is near unity in the case of ``sine" data given
by Eq. (\ref{2dSine}), this problem does not arise for $\sigma = 1$
for ``sine" data. Similar comments apply, of course, in the 3D case
of equation (\ref{NLPoisson}).
\label{footnote}
}

\begin{eqnarray}
         r_{GS} & = & h^{-2} \left( 
                            \tilde{u}_{i+1,j} + \tilde{u}_{i-1,j}
                          + \tilde{u}_{i,j+1} + \tilde{u}_{i,j-1}
                          - 4 \tilde{u}_{i,j} \right) \cr
           &\mbox{}& + \sigma  \tilde{u}^2_{i,j} -
                          f_{i,j}
\end{eqnarray}

\begin{equation}
           \tilde{u}^{\rm new}_{i,j} = \tilde{u}^{\rm old}_{i,j} 
                             - { r_{GS} \over
                             2 \sigma \tilde{u}_{i,j} - 4h^{-2}}.
\label{2dgaussSeidel}
\end{equation}

\hfil\break
\noindent{\em Restriction and Prolongation Operators}\hfil\break

\begin{figure}
\centering
\centerline{\includegraphics[width=5.2in,height=1.23in]{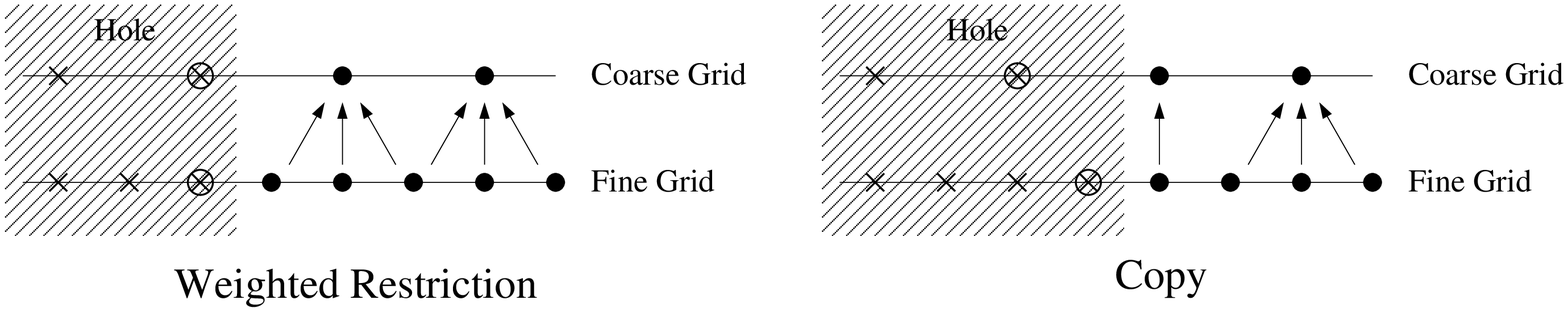}}
\caption{1-D schematic of scheme for inner boundary and 
restriction scheme.  
The circles on the rightmost X's indicate that this is where the 
Dirichlet conditions are applied, i.e., there {\em are} data on these
points for all quantities except $\myL^h \tilde{u}^h$ (which
appears in (\ref{tauh2hdef})), which we do not compute on the
boundary.
One {\em could} use these points in weighted restriction even 
in the case shown in the right panel.  However, we choose never to use
these boundary data in weighted restriction, and instead do a 
simple ``copy" operation.
}
\label{fig_restrict_scheme}
\end{figure}

The restriction operator $\restrict$ we use is the so-called
``half-weighted'' average on normal interior points, in which coarse grid
values (indexed by $I$ and $J$ for clarity) are a weighted average
of the fine grid values over a nearby region of the physical domain:
\begin{equation}
            \tilde{u}^{2h}_{I,J} = \restrict \tilde{u}^h = 
               {1\over 2} \tilde{u}^h_{i,j} + 
               {1\over 8} \left[ 
                    \tilde{u}^h_{i+1,j} + \tilde{u}^h_{i-1,j} +
                    \tilde{u}^h_{i,j+1} + \tilde{u}^h_{i,j-1} \right],
\label{2drestrict}
\end{equation}
where $i = 2I - 1$ and $j = 2J -1$.
Along the outer boundary, we perform a simple copy operation, 
$\tilde{u}^{2h}_{I,J} = \tilde{u}^h_{i,j}$.
We use (\ref{2drestrict}) for all restriction operations on interior 
grid points with one exception: 
We only use weighted restriction if none of the fine 
grid points used in the restriction operator is on or inside the
boundary of the excised region;  otherwise we use a
``simple injection" or ``copy" operation.  
This is illustrated in Figure \ref{fig_restrict_scheme}.
In conjunction with our definition of the inner boundary location, 
this use of a ``copy'' operation near the inner boundary
is the central insight for preserving second 
order accuracy near the excision region.

For the prolongation operator $\prolong$, we use simple bilinear 
interpolation.

\subsubsection{2D Results}\hfil\break

\begin{figure}
\centering
\centerline{\hspace{3cm}\includegraphics[width=3.5in]{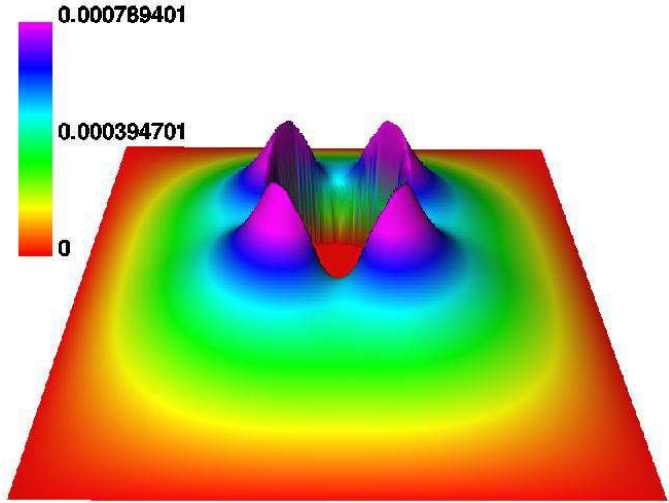}\hspace{0cm}}
\vspace{-2.3cm}
\caption{
A plot of the solution error $e = u - \tilde{u}^h$ for a system
physical parameters $\sigma=1$ and $l_x=l_y=1$.  We use a central
circular hole of radius $r_{\rm mask} = 0.129$,
where this number is chosen such that the excision masks
on different grid levels do {\em not} match up with one another,
i.e., to make the solution {\em harder} to obtain, in order to
demonstrate the robustness of our multigrid scheme.  We see that
the error is largest and sharpest next to the circular hole in the
center.  Thus, concentrating our smoothing operations near the hole
is likely to increase the efficiency of the solver.  These data
are obtained from a run with parameters $L_{\rm max}=8$, $L_{\rm
min}=2$, 1 V-cycle and 2 pre- and 2 post-CGC smoothing sweeps.
} 
\label{fig_3dplot}
\end{figure}

One of the first things we notice when
graphing preliminary numerical solutions is that the solution error 
$e\equiv u - \tilde{u}^h$ seems to be largest and ``most in need of
smoothing'' (i.e., having high-frequency components) in the immediate
vicinity of the excision region, as shown in Figure \ref{fig_3dplot}.

\begin{figure}
\centering
\centerline{
{\hskip -4.6cm} 
\vbox{\includegraphics[width=2.2in]{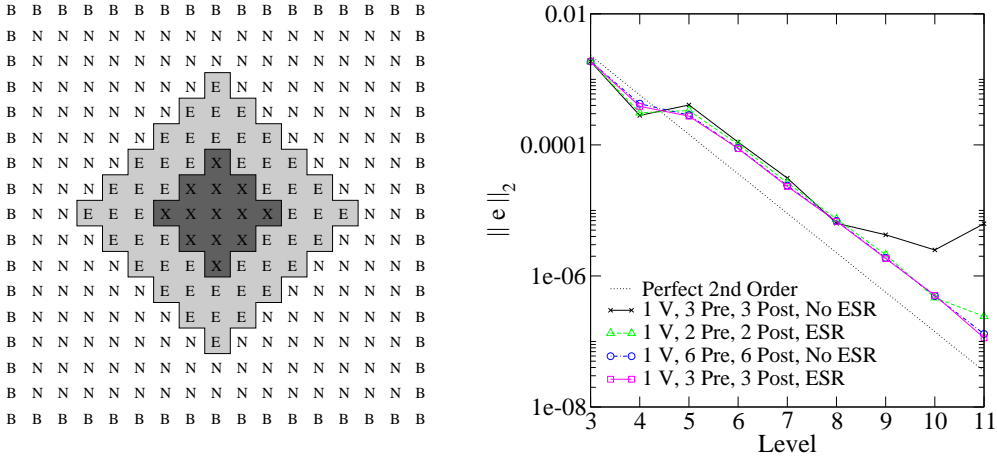} \vskip 0.4cm}
{\hskip -3.3cm}
\vbox{ \vskip -0.5cm}
\includegraphics[width=2.67in]{esr_vs_noesr.eps}  
{\hskip -1cm} }
\caption{
Left panel: Schematic of a ${\rm level} =4 grid$ ($2^{\rm level}+1$
grid points per side) showing excised points (X's), normal interior
points (N's), boundary points (B's) and the Extra Smoothing Region
(ESR, E's).  In this case, the excision region has radius $r_{\rm
mask}=0.129$.  Right panel: Convergence behavior of 
Eqs. (\ref{2dPoisson})-(\ref{2dgaussSeidel}) with and without
the use of the ESR.  Here we show the L2 norm of the solution error
$e = u - \tilde{u}$.  We assign the width of ESR to be ${\rm
level}-1$ grid points on either side of the hole, and we smooth
twice as often over the ESR as over the normal interior points.
Since the error is concentrated near the hole, we see that using
2 pre- and post-CGC smooths with the ESR (i.e., using 4 smooths in
the ESR, and 2 smooths elsewhere) offers a substantial improvement
over the use of 3 pre- and post-CGC smooths over the whole domain.
We see that at high resolutions (${\rm level} = 9$ through $11$)
the error begins to rise again.  To obtain results in keeping with
second order convergence at resolutions up to ${\rm level} = 11$,
more work is required.  The use of 3 pre- and post-CGC smooths {\em
with} the ESR (i.e., 6 smooths in the ESR, and 3 elsewhere) produces
results comparable to those obtained by using 6 pre- and post-CGC
smooths over the whole domain, however the former requires a fraction
of the computational cost of the latter.  
} \label{fig_esr}
\end{figure}

Thus we may perform only a few smooths on the entire grid and apply
extra smoothing runs in an ``extra smoothing region'' (ESR) around
the excision region, such as that shown in in the left panel of
Figure \ref{fig_esr}.  In our implementation, we choose the width of the ESR
to be a number of grid points which is one less than the current
multigrid level number (e.g.,  on a level 4 grid, the ESR has a
width of 3 grid points), and we smooth over the ESR twice as often as
over the rest of the domain.  We can compare the results of using the
extra smoothing region to the results of smoothing over the entire
domain.  Such a comparison is shown in the right panel of Figure
\ref{fig_esr}.  In our multigrid implementation, we always do twice
as many smooths in the ESR as in the rest of the interior.

\hfil\break
\noindent{\it Two Holes}\hfil\break

One of our principal applications for the multigrid solver will be
to solve the constraint equations for binary black hole spacetimes.
Thus we do a check to make sure that, at least for the nonlinear
Poisson equation, our general method properly handles domains with
multiple holes.  One such case is shown in Figure 5.

\begin{figure}
\centering
\centerline{
{\hskip -4.6cm} 
\vbox{\includegraphics[width=2.2in]{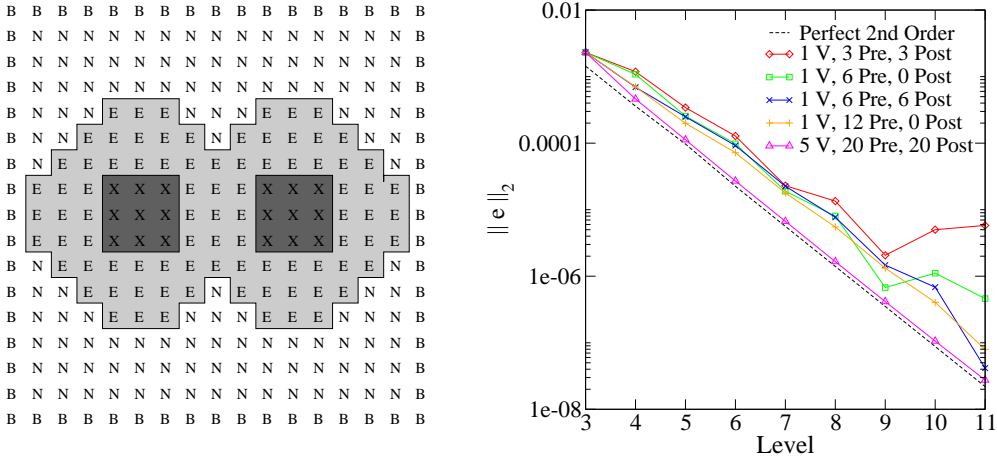} \vskip 0.4cm}
{\hskip -3.3cm}
\vbox{ \vskip -0.5cm}
\includegraphics[width=2.67in]{twoholes_pre_vs_post.eps}  
{\hskip -1cm} }
\caption{
Comparison of pre-CGC smoothing vs. post-CGC smoothing, for a domain
with two circular holes of radius $r_{\rm mask}$=0.1, separated by a 
distance $d=0.2$, again for Eqs. (\ref{2dPoisson})-(\ref{2dgaussSeidel}).  
Since a pre-CGC smooth requires the same number of operations
as a post-CGC smooth, runs with 1 V-cycle, 3 pre- and 3 post-CGC smooths 
are as costly as runs with 1 V-cycle, 6 pre- and 0 post-CGC smooths.
However the latter parameters yield a somewhat lower error at all resolutions.
In another case, we compare 6 pre- and 6-post CGC smooths with 
12 pre- and 0 post-CGC smooths, with the similar result that
12 pre- and 0 post-CGC smooths yield lower error.   
We note that
the general shapes of these graphs seem to alternate between
`turning down' and `turning up', with the cases involving more 
computational cost producing results closer to the ideal line.  This
suggests that, for a given range of resolutions, one is able to 
obtain results closer and closer to second order convergence as one
increases the computational cost; an `extreme' case of 5 V-cycles, with
20 pre- and 20 post-CGC smooths is shown to illustrate this.   
It is important to note that post-CGC smoothing is not negligible 
in all cases: One needs post-CGC
smooths particularly for runs involving multiple V-cycles in order
to ensure convergence, e.g., a 3 V-cycle run with 12 pre- and 0 post-CGC
smooths {\em immediately} begins to `blow up' as the resolution is 
increased.  Adding post-CGC smooths to such a run considerably 
improves the convergence.
\label{fig_pre_vs_post}
} \end{figure}

There is a contrast between the relative importance of pre-CGC
smoothing versus post-CGC smoothing.  Pre-CGC smoothing prepares
the data in such a way that restriction will not produce aliasing
errors, whereas post-CGC smoothing in large part corrects for
errors brought in via the use of (linear) interpolation.  It seems
plausible that the former case (pre-CGC) would require at least as much
smoothing as the latter (post-CGC); we see evidence for this in
the two-hole solutions, such as those used to produce Figure
\ref{fig_pre_vs_post}.  Again we are able to achieve second order
convergence, for instance with 1 V-cycle, 12 pre-CGC sweeps and no
post-CGC sweeps, and with more effort: 5 V-cycles, 20 pre- and post-CGC
sweeps, shown for comparison as a ``perfect'' multigrid result.

\subsection{Three dimensions}
For solving equations in 3D, we proceeded by steps, performing a few
test cases until arriving at a code which solves an equation which is
similar to the Hamiltonian constraint in 3+1 general relativity.

The first test case is similar to the 2D case discussed above, i.e., we 
choose $f(x,y,z)$ such that the exact
solution to (\ref{NLPoisson}) is
\mkeq{ u(x,y,z) = \sin(\pi l_x x) \sin(\pi l_y y) \sin(\pi l_z z)
\label{3DSinusoidalCase}}
where $l_x$, $l_y$ and $l_z$ are integers which we set to unity. 
We excised a sphere of radius $r_{\rm mask}=0.129$, which is a value
chosen simply to ensure that the size of the mask is different on
different grid levels, i.e. to make the test a bit more difficult
than if we had chosen some multiple of the mesh spacing.
For the test case involving a sinusoidal solution 
(\ref{3DSinusoidalCase}), the results are very similar to those of
the 2D case.
Having verified that the multigrid solver could adequately handle such
a system and achieve second order accuracy, we proceeded to the next step.

The next test case involves an equation (\ref{ConstrEq})
which corresponds to the solution of Hamiltonian constraint equation
for conformally flat spacetimes system\cite{YorkPiran}:
\begin{equation}
   \nabla^2 u - K^2 u^5 + {A^2 \over u^7} = f, 
\label{ConstrEqAgain}
\end{equation}
where $K^2$ and $A^2$ are arbitrary positive real constants related to
the expansion rate of the 3-space, and $r$ 
is the usual radial coordinate.  
We adjust the matter source term $f(x,y,z)$
such that the continuum solution is
\mkeq{ u(x,y,z) = 1 + {2 M / r}. \label{3DRobinCase} }
Here $r$ is the usual radial coordinate, and we choose $M=1$.
The excision region contains $r=0$ and the function values in
this region are not calculated.  
On the inner boundary (the boundary of the excision region), we use
values of the continuum solution as Dirichlet conditions on 
{\em only the finest grid}, and on coarser grids, we use 
data copied or extrapolated from finer grid data.
 One can simply copy values
from the fine grid to the coarse grid wherever the boundary points
on the fine and coarse grids coincide.   Wherever they do not
coincide, i.e., wherever the coarse grid boundary points are `interior'
to the fine grid boundary points, we find it quite adequate to use
quadratic extrapolation (in one dimension) from the neighboring
(non-excised) fine grid points to obtain the value at the location
of a coarse grid boundary point.  The results we obtain appear to
be rather insensitive to the form of the extrapolation used (e.g.,
the shape of the stencil).  The use of extrapolated fine-grid data
is an important difference
from the way inner boundary values are obtained in the 2D code, and
does constitute a great improvement over the generic applicability 
of our method.

For the outer boundary in this case, we can either apply Dirichlet 
conditions using the continuum solution $u(x,y,z)$, 
or we can
use a mixed boundary condition called the {\em Robin} condition.
The Robin condition requires 
\begin{equation}
{\partial \over \partial r} [ r (u-1) ] = 0,
\label{Robin}
\end{equation}
on the boundary.
The Robin condition is the standard condition used in relativity when 
a solution with the form (\ref{3DRobinCase}) is desired
(see, e.g., \cite{YorkPiran}).
However $M$ is not a known constant until the solution is known,
so the condition which the solution (\ref{3DRobinCase})
satisfies, namely (\ref{Robin}), is implemented as the boundary condition.

There are several ways to implement the Robin condition.
Rather than taking derivatives in the radial direction as required
by (\ref{Robin}), we follow Alcubierre \cite{Miguelcomment}
and instead take derivatives only in directions normal to the faces 
of our cubical domain. This approach is much simpler 
to implement than using radial derivatives.
We implement this using first order difference stencils;
for the $+x$ face of the domain, at location $x_i$, the stencil is
simply
\begin{equation}
   \tilde{u}_{i,j,k} = 1 + (\tilde{u}_{i-1,j,k} - 1) 
       { r_{i-1,j,k} \over  r_{i,j,k}  },
\label{firstordercopy}
\end{equation}
where $r_{i,j,k}$ is the radial distance to lattice location $(i,j,k)$.
We have also implemented alternative stencils:
second order stencils in the normal direction,
and a first order stencil in the radial direction.


\hfil\break
\noindent{\em Smoothing Operations}\hfil\break

We use a simple ``red-black'' 
Gauss-Seidel Newton iteration to smooth the error, i.e.,
\begin{eqnarray}
         r_{GS} & = & h^{-2} \left( 
                            \tilde{u}_{i+1,j,k} + \tilde{u}_{i-1,j,k}
                          + \tilde{u}_{i,j+1,k} + \tilde{u}_{i,j-1,k}
                          \right. \cr
    &\mbox{}&  \hspace{0.5cm} \left.  
                          + \tilde{u}_{i,j,k+1} + \tilde{u}_{i,j,k-1}
                          - 6 \tilde{u}_{i,j,k} \right) \cr
           &\mbox{}& - K^2  \tilde{u}_{i,j,k}^5 
                     + A^2  \tilde{u}_{i,j,k}^{-7} -
                          f_{i,j,k}
\end{eqnarray}

\begin{equation}
           \tilde{u}_{i,j,k} = \tilde{u}_{i,j,k} - { r_{GS} \over
       -5 K^2 \tilde{u}_{i,j,k}^4 - 7 A^2 \tilde{u}_{i,j,k}^{-8} - 6h^{-2}} 
\end{equation}

\hfil\break
\noindent{\em Restriction and Prolongation Operators}\hfil\break

For the restriction operator $\restrict$, we use a weighted average 
which is defined 
via a $3\times 3 \times 3$ stencil, with a weight of 1/8 on the central
point, 1/16 on the center of each face, 1/32 on the center of each edge
and 1/64 on each corner, i.e.:
\cite{BAM}
\begin{eqnarray}
\tilde{u}^{2h}_{I,J,K} & = &  \restrict \tilde{u}^h = 
         {1\over 8} \left[ \tilde{u}^h_{i,j,k} \right. \cr
    &\mbox{}& + {1\over 2} \left(
           \tilde{u}^h_{i+1,j,k} + \tilde{u}^h_{i-1,j,k}
         + \tilde{u}^h_{i,j+1,k}  \right. \cr
    &\mbox{}&  \hspace{0.6cm} \left.  
         + \tilde{u}^h_{i,j-1,k}
         + \tilde{u}^h_{i,j,k+1} + \tilde{u}^h_{i,j,k-1}
         \right)   \cr
     &\mbox{}&    + {1\over 4}\left( 
            \tilde{u}^h_{i,j+1,k+1} + \tilde{u}^h_{i,j-1,k+1}
         + \tilde{u}^h_{i,j+1,k-1} + \tilde{u}^h_{i,j-1,k-1} \right. \cr
    &\mbox{}&  \hspace{0.6cm} \left.  
         + \tilde{u}^h_{i+1,j,k+1} + \tilde{u}^h_{i-1,j,k+1} 
         + \tilde{u}^h_{i+1,j,k-1} + \tilde{u}^h_{i-1,j,k-1} \right. \cr
    &\mbox{}&  \hspace{0.6cm} \left.  
         + \tilde{u}^h_{i+1,j+1,k} + \tilde{u}^h_{i-1,j+1,k}
         + \tilde{u}^h_{i+1,j-1,k} + \tilde{u}^h_{i-1,j-1,k}
         \right)  \cr
   &\mbox{}&  + {1\over 8} \left(
            \tilde{u}^h_{i+1,j+1,k+1} + \tilde{u}^h_{i-1,j+1,k+1}
         + \tilde{u}^h_{i+1,j-1,k+1} \right.\cr
    &\mbox{}&  \hspace{0.6cm}  \left.  
         + \tilde{u}^h_{i-1,j-1,k+1} + \tilde{u}^h_{i+1,j+1,k-1} + \tilde{u}^h_{i-1,j+1,k-1} \right. \cr
    &\mbox{}&  \hspace{0.6cm} \left. \left.  
         + \tilde{u}^h_{i+1,j-1,k-1} + \tilde{u}^h_{i-1,j-1,k-1}
                             \right)
                           \right]
\end{eqnarray}
As in the 2D case, if any of the
find grid points included in the weighting are inner boundary points,
then we use a simple copy instead of weighted restriction 

For the prolongation operator $\prolong$, 
we use simple trilinear interpolation.

\subsection{3D Results}
\begin{figure}
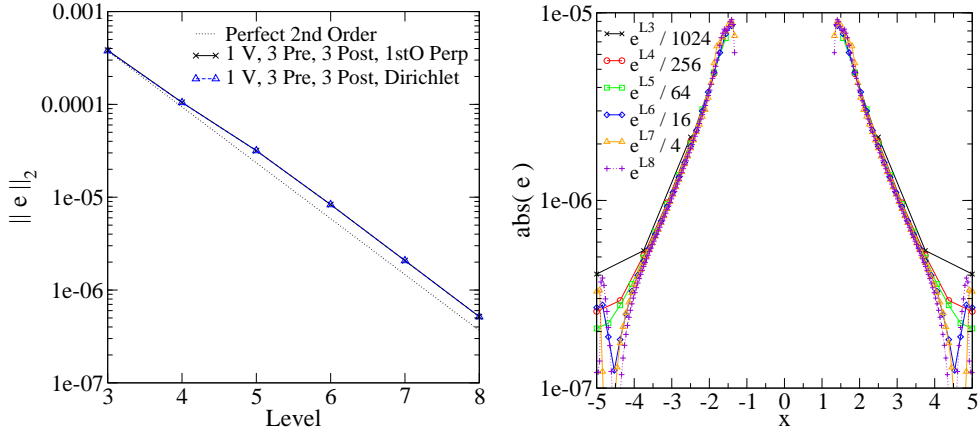

\centering
\centerline{
\includegraphics[width=2.5in]{robin.eps}
\hspace{0.1cm}
\includegraphics[width=2.45in]{robin2.eps}
}
\caption{
Convergence results for 3D solutions to (\ref{ConstrEqAgain}) of the form
$u = 1 + 2 M /r$, for runs
in which $A^2=K^2=1$, $l_{\rm min}=3$, and $r_{\rm mask}=1.29$, 
for a domain $[-5,-5,-5]$ to $[5,5,5]$.
Left panel: 
A logarithmic plot of the L2 norm of the solution error  
$e = u - \tilde{u}^h$, showing a comparison between outer boundary 
conditions.
Using the ``first order perpendicular" (FOP) implementation 
(\ref{firstordercopy}) of the Robin boundary condition (\ref{Robin})),
we obtain convergence results which lie on top of those obtained using
a Dirichlet outer boundary condition (where the values of the continuum
solution are supplied at the outer boundary).  These results also
run parallel to the line for perfect second order behavior.  
Right Panel: A logarithmic plot of $e$ itself, at the end of each 
V-cycle in the
Full Multigrid Algorithm.  These results correspond to
the ``1V, 3 Pre, 3 Post, 1st order Perp" case shown in the left
panel.  Here we show a 1D slice along the x-axis, and we have
divided coarser grid values by appropriate powers of four in order
to make the comparison.  We see second order convergence over the
interior of the domain and, importantly, in the vicinity of the excision
region.
Near the outer boundary, the magnitude of the error is roughly 
second order (thus it does not noticeably affect the graph shown 
in the left panel), however its shape is resolution-dependent.
This feature may arise from the use of the FOP condition. 
}
\label{fig_robin}
\end{figure}

A graph of the convergence behavior for the solution of the
conformally flat background Hamiltonian constraint, 
Eq.  (\ref{ConstrEqAgain}), is given in Figure \ref{fig_robin}.  
The left panel shows
that we obtain lines which run parallel to the line for second order
behavior at high resolutions, for both a Dirichlet outer boundary
condition and the ``first order perpendicular'' implementation of
the Robin condition suggested by Alcubierre \cite{Miguelcomment}.
In fact, the two graphs we obtain lie on top of each other.
The right panel shows the convergent solution error of our
solutions (i.e. the difference between our solutions and the exact, 
continuum solution)

\section{Toward More General Applications}
We have made use of a known exact solution in two key elements of
this paper: (1) supplying values for Dirichlet conditions on the
inner boundary, and (2) calculating the solution error 
$e \equiv u - \tilde{u}^h$ for measuring the accuracy and convergence
of the code.  Since we wish to use the multigrid solver for situations
in which an exact solution is not known across the whole domain
(such as in a solution for the conformal factor in a general binary
black hole spacetime), we describe here the modifications to the
previous discussion in the absence of an exact solution.

In the 2D code we used the continuum solution to supply Dirichlet 
conditions on the inner boundaries of all multigrid levels, however
for the 3D code we only used the continuum solution on the finest
grid, and then extrapolated or copied data from finer grids to coarser
grids.  (There was nothing about the 2D case that made extrapolation
impossible; rather it was simply a later feature which was added
to the more advanced 3D code.) Although we 
were able to employ this extrapolation effectively 
in the cases we tried
--- using a very
simple, almost arbitrary, 1D extrapolation method --- it may be that
such a method would be unstable for certain classes of equations.
Although we see no evidence for this, and given the generality of the
multigrid method one may expect much of what is found in this paper
to apply to other systems of interest.  We cannot offer
complete assurance that extrapolation will work for all elliptic systems.


The assumption that an exact value is known at points on the finest
grid may not be appropriate for many of the solutions of interest
to numerical relativists.   One may encounter systems with 
Neumann or Robin conditions at the inner boundary.
Although we have not considered such cases explicitly, we suggest 
straightforward differencing similar to the treatment of the Robin
conditions we reported above.  

\begin{figure}
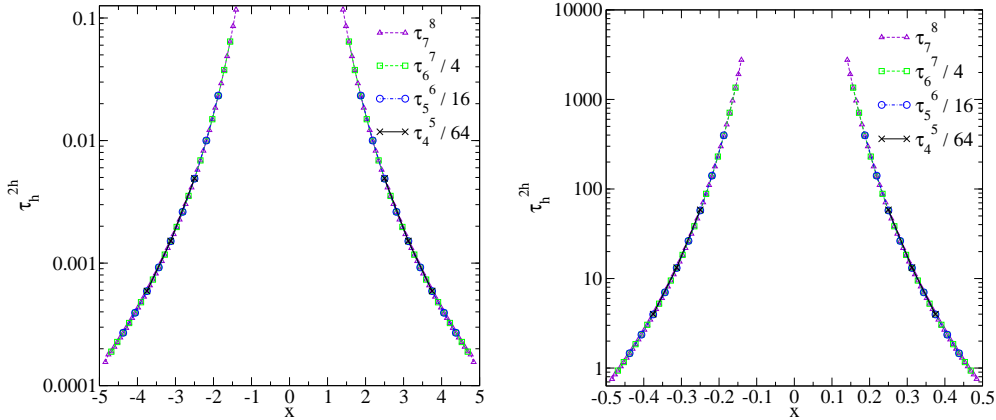

\centering
\centerline{ 
   \includegraphics[width=2.5in]{logabstauh2h_xmax5,0.eps}
   \hspace{0.2cm}
   \includegraphics[width=2.5in]{logabstauh2h_xmax0,5.eps}
}
\caption{
A 1D slice through the center of a 3D domain, for the truncation error 
of a solution to Eq. (\ref{ConstrEqAgain}) with Robin outer
boundary conditions, seeking a solution of the form $u = 1 + 2 M / r$.
The run parameters $K^2$, $A^2$ and $M$ are set to unity, and we perform
1 V-cycle, 3 pre-CGC smooths and 3 post-CGC smooths.
Left panel: On the domain $[-5,-5,-5]$ to $[5,5,5]$, with
$r_{mask}=1.29$.
Right panel: On the domain $[-0.5,-0.5,-0.5]$ to $[0.5,0.5,0.5]$, with
$r_{mask}=0.129$.  (This smaller domain is intended to provide 
a stronger test of the algorithm than the larger domain.)
The notation $\tau_x^y$ in the figure legends denotes which grid levels
were used to calculated $\tau_h^{2h}$ for a given set of points.
We see that, on both domains, 
the $\mathcal{O}(h^2)$ error function is independent of resolution,
in accordance with the idea of Richardson extrapolation.
Since the continuum solution is $u(x,y,z) = 1 + 2 M / r$, 
we expect the error function to tend toward large values as 
$r\rightarrow 0$.
In these graphs, we have cut out the values of $\tau_h^{2h}$ immediately
adjacent to the inner boundary.
}
\label{fig_tauh2h}
\end{figure}

When one cannot measure the accuracy of the code by calculating the
solution error, one can still gain a measure of the convergence of
the code by monitoring the relative truncation error $\tau_h^{2h}$.
One should disregard values at points adjacent to inner boundary points,
as these tend to be poorly defined and exhibit blow-up behavior,
but all other points are eligible for comparison.
A plot of $\tau_h^{2h}$ at various grid levels for a 3D solution is 
shown in Figure \ref{fig_tauh2h}, demonstrating the strict second order
behaviour of the solution on two different domains.

\section{Conclusions}
We can offer five tips  for those who wish to implement
2D or 3D multigrid methods for domains with holes.
\begin{itemize}
\item $1)$ The first and most important tip is really twofold:
\begin{itemize}
\item apply inner boundary conditions on points immediately {\em
interior to} (rather than exterior to, as is often done) some
spatial surface such as a circle or sphere of a given radius.  This
has the effect that the extent of excision region is smaller on
coarser grids than on finer grids.
\item when performing restriction
operations, use weighted restrictions on all interior
points, {\it except} in cases where the weighted operator would include
points where the inner boundary conditions are applied.  For these
latter cases, use a simple copy operation.
\end{itemize}
\item $2)$ Concentrate smoothing operations
where they are most needed.  We have done this by arbitrarily
defining an ``extra smoothing region'' around any hole, and
performing twice as many smooths in this region as in the rest of
the domain.  One can imagine more sophisticated schemes, which look
at derivatives of the solution or the local truncation error as
indicators of where the extra smoothing should be performed and
how much extra work should be performed there.

\item $3)$ It may be the case that pre-CGC smoothing is much
more effective than post-CGC smoothing for a given problem,
particularly if only one V-cycle is performed.  Concentrating the
bulk of the smoothing operations into the pre-CGC smooths
results in a faster route to the solution at a desired accuracy.

\item $4)$ This repeats the advice of Alcubierre \cite{Miguelcomment}
(Radial) Robin boundary conditions can be adequately mimicked by
applying them only in the normal directions, thus simplifying the
computer code and reducing the computation effort slightly.
The Robin conditions should be
applied at all grid levels; i.e. it is {\it not} the case that one should
apply Robin conditions on the finest grid only and then use those
values as Dirichlet conditions on coarser grids.

\item $5)$ This tip is intended for use in situations where
an exact solution is not available; we have less experience with
these situations. However, we find that a naive quadratic
extrapolation from nearby points is adequate to provide values for
Dirichlet boundary conditions on the edge of the hole. Further
(more technically), when monitoring the relative truncation
error $\tau_h^{2h}$, points immediately adjacent to inner boundary
points have erratic behavior, and should not be considered in judging
the behavior of the solution.

\end{itemize}

We clearly have some further work ahead in order to consider more
general systems in which our inner boundary data is not so easily
obtained, and we also need to consider systems more complex than
the Hamiltonian constraint for a conformally flat background, as 
treated in this paper.  It is thus
our hope that the techniques described here will also be of relevance
for more interesting systems of elliptic equations.  We are developing
a solver for the full set of constraint 
equations (testing it using the Kerr-Schild-type binary black hole
data described in \cite{Bonning}), and see promising results 
using precisely the techniques given in this paper.
Further improvements in the speed of multigrid scheme, such as the use
parallelism and adaptivity \cite{AdaptiveMG,DavidBrown,Baker} will
also likely be necessary for its practical application in a
constrained evolution code.

The 2D and 3D codes described here are available for use by the 
community, by download from
\verb+http://wwwrel.ph.utexas.edu/~shawley/mg_ex.html+.
We encourage other users of our approach. However, the codes are 
``as is", and we cannot maintain them, nor correct problems that 
may arise from their use.

\section{Acknowledgments}
We thank Matthew Choptuik for helpful discussions, and for the use of
his notes and his 2D FAS multigrid code.  We also thank Miguel
Alcubierre, Bernd Br\"ugmann, Gregory Cook, Lee Lindblom, Mark
Scheel, Frans Pretorius, James Isenberg and Sergio Dain for helpful 
discussions.
We thank Jonathan Thornburg and Ian Hawke for their helpful comments
on a previous draft of this paper, and we thank Roberto Gomez for 
pointing out the Udaykumar et al. reference \cite{Udaykumar} to us.  
This work was supported by NSF grant PHY
0102204.  Portions of this work were conducted as part of the Caltech
Visitors Program on the Initial Data Problem, and at the Kavli
Institute for Theoretical Physics, The University of California at
Santa Barbara, under NSF grant PHY 9907947.

\end{document}